\documentclass[runningheads]{svmult}

\usepackage{makeidx}   
\usepackage{graphicx}  
\usepackage{subeqnar}  
\usepackage{multicol}  
\usepackage{physprbb}  
\makeindex             

\begin{document}
\title*{Measuring Galaxy Disk Mass with the
\protect\newline
SparsePak Integral Field Unit on WIYN}
\toctitle{Measuring Galaxy Disk Mass with the
\protect\newline 
SparsePak Integral Field Unit on WIYN}
%
%
\titlerunning{SparsePak measurements of Galaxy Disk Mass}
%
\author{Marc Verheijen\inst{1}
\and Matthew Bershady\inst{1}
\and David Andersen\inst{2}}
\authorrunning{Marc Verheijen et al.}
%
%
\institute{University of Wisconsin, Madison WI 53706, USA
\and Max Planck Institute for Astronomy, Heidelberg, Germany}

\maketitle              


%

\vspace{8mm}

We present first results from the commissioning data of the SparsePak
Integral Field Unit on the WIYN telescope. SparsePak is a bundle of 82
fibers, arranged in a sparsely filled, 76$\times$77 arcsec hexagonal
grid. It pipes light from one of the Nysmith foci to the Bench
Spectrograph. See http://www.astro.wisc.edu/
\~\protect{m}ab/research/sparsepak and Bershady et al. (2002a) for
more details. The fibers are each 5 arcsec in diameter and are thus
suitable to obtain spectroscopy on extended objects of lower surface
brightness such as galaxy disks. In fact, SparsePak was purposely
developed to measure stellar kinematics and velocity dispersions in
disk galaxies with the ultimate goal of constraining the surface
densities and masses of stellar disks. Such a measurement is crucial
to improve our understanding of the principal dynamical components in
disk galaxies from rotation curve decompositions.

As part of the SparsePak commissioning program we observed several
galaxies, one of which was NGC 3982, a blue and very high surface
brightness galaxy with $\mu_0^{\rm obs}$(B)=19.3 mag/arcsec$^2$ and
B$-$K$^\prime$=3.4 (Tully et al. 1996). It was mapped in three
spectral regions; around 5130\AA (MgI), 6680\AA (H$\alpha$) and
8660\AA (CaII-triplet) with FWHM spectral resolutions of 24, 16 and 37
km/s respectively.  About 3 dozen template stars with a range in
T$_{\rm eff}$ and surface gravity were observed as well by drifting
them across many fibers.

Three pointings were taken in the H$\alpha$ line to fill the hexagonal
grid. This yielded a contiguous H$\alpha$ velocity field of very high
signal-to-noise. Fitting a tilted-ring model to this velocity field
yielded a kinematic inclination of 26$\pm$2 degrees, consistent with
the isophotal ellipticity, and an H$\alpha$ rotation curve at 5 arcsec
resolution which supplements a low resolution HI rotation curve.

Stellar absorption line observations were taken with a single
SparsePak pointing, aimed at measuring the vertical velocity
dispersion of the stars. Here we focus on the stellar velocity
dispersions measured from the deepest CaII line. In the inner regions
of the galaxy, the signal-to-noise in the spectra of individual fibers
is high enough to determine the line centroids and to construct a
stellar velocity field. Comparison of the H$\alpha$ and stellar
velocity fields clearly shows the effects of asymmetric drift inside
1.5 disk scale lengths h$_{\rm R}$.  Outside this radius the
asymmetric drift approaches zero.

The layout of the fibers and the near face-on orientation of the
galaxy allows for azimuthal averaging of 6, 6, 6, 12 and 18 fibers
in 5 annuli to improve the signal-to-noise in the CaII absorption
line in the outer regions of the galaxy. The outermost annulus, with 18
fibers, has a radius of 3.5 h$_{\rm R}$. Before averaging, the
projected rotational velocities were taken out, using the centroids of
the high signal-to-noise H$\alpha$ lines. Stellar velocity dispersions
were determined by convolving the spectrum of a K0.5-III template star
with a Gaussian of varying FWHM and finding the best match to
the azimuthally avaraged galaxy spectra.

The results are summarized in Figure 1 and in Bershady et
al. (2002b). The plotted velocity dispersions are the $\sigma$ of the
convolution Gaussian. No corrections were applied for contributions
from the projected radial and tangential velocity dispersion
components which are estimated to contribute less than 13\%.  The
measured velocity dispersion is fairly constant with radius except for
the outermost point where a significantly lower velocity dispersion is
observed. Assuming an isothermal (sech$^2$) vertical density profile,
a mass surface density can be derived for a variety of scale heights
based on results from recent work by Kregel et al. (2002). This mass
surface density includes any non-stellar component in the disk and is
a factor $\sim$3.5 higher than the surface density of the Milky Way in
the solar neighborhood (Kuijken \& Gilmore 1991), indicated by the
short dashed line in the middle panel. Given the K$^\prime$ luminosity
profile, the mass-to-light ratio of the disk can be computed as a
function of radius. A weighted radial average of
(M/L$_{K^\prime}$)=0.18 is used in a rotation curve decomposition
shown in Figure 2 which implies a substantially ($\sim 25$\%)
sub-maximum, yet very high SB disk.

\begin{figure}[t]
\begin{minipage}[b]{65mm}
\includegraphics[width=\textwidth]{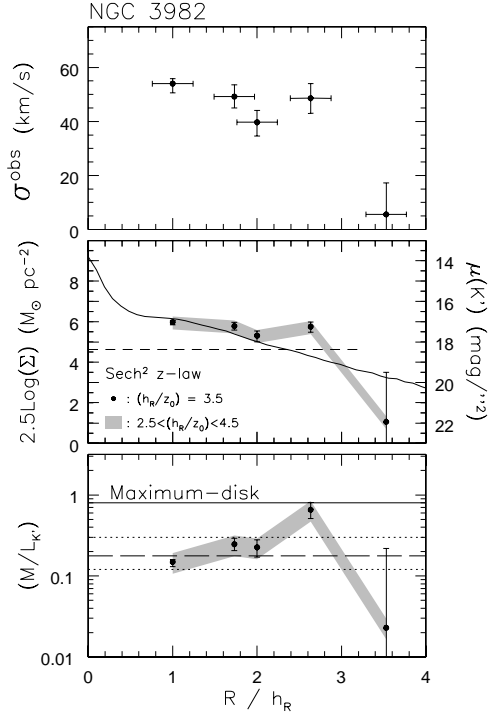}
\end{minipage}
\hspace*{\fill}
\begin{minipage}[b]{45mm}
\caption[1]{Upper panel: Azimuthally averaged stellar velocity
dispersions, corrected for instrumental and template broadening.
Horizontal bars indicate the radial bin widths. Note the sudden drop
beyond 3 disk scale lengths. Middle panel: Points indicate the
inferred mass surface density for an isothermal vertical density
profile and a range of scale heights. The solid line shows the radial
K$^\prime$ surface brightness profile. Bottom panel: derived M/L of
the disk in the K$^\prime$ band. The solid line shows the maximum disk
value allowed by the rotation curve. The dashed line indicates the
weighted radial average value and the dotted lines relate to the
dotted lines in the right panel of Figure 2.}
\end{minipage}
\label{eps1}
\end{figure}

\begin{figure}[t]
\begin{center}
\includegraphics[width=\textwidth]{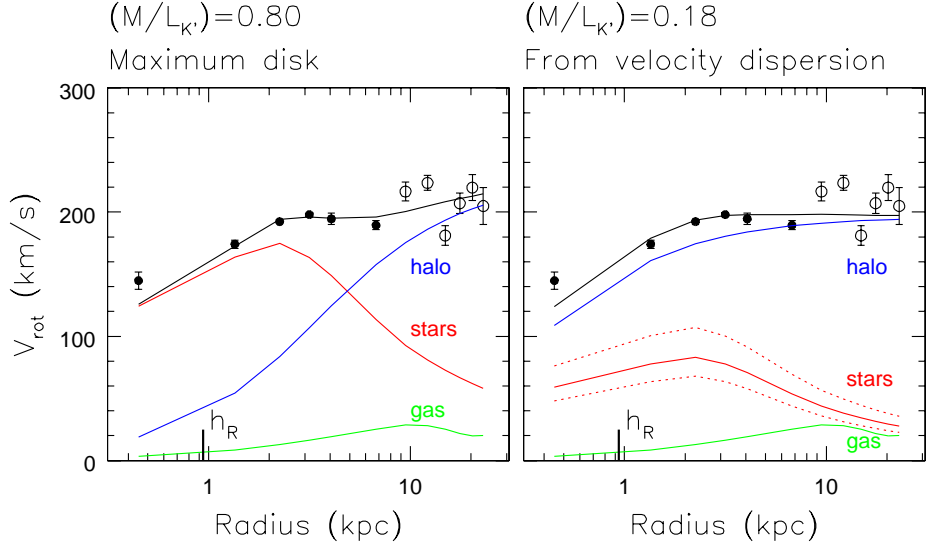}
\end{center}
\caption[2]{Rotation curve decompositions for NGC 3982. Solid points
are from SparsePak, open points are from HI measurements. Left panel:
maximum-disk decomposition which implies a stellar mass-to-light ratio
of 0.80 in the K$^\prime$ band. Right panel: rotation curve
decomposition with a disk M/L of 0.18 based on the stellar velocity
dispersion measurement of SparsePak. This suggests that NGC 3982 has a
strongly sub-maximum disk. This is surprising given the fact that the
stellar disk has a very short scale length and and extremely high
surface brightness.}
\label{eps2}
\end{figure}

This work was supported by NSF grant AST-9970780.

%

\end{document}